\documentclass[twocolumn,aps,showpacs,amsmath,amssymb]{revtex4}

\usepackage{graphicx}
\usepackage{dcolumn}
\usepackage{bm}

\begin{document}

\title{
Nucleus-nucleus potential with shell-correction contribution
}%

\author{
V. Yu. Denisov
}%

\affiliation{%
Institute for Nuclear Research, Prospect Nauki 47,
03680 Kiev, Ukraine \\
Faculty of Physics, Taras Shevchenko National University of Kiev, Prospect Glushkova 2, 03022 Kiev, Ukraine
}%

\date{\today}

\begin{abstract}
The full relaxed-density potential between spherical nuclei is considered as a sum of the macroscopic and shell-correction contributions. The macroscopic part of the potential is related to a nucleus-nucleus potential obtained in the framework of the extended Thomas-Fermi approach with the Skyrme and Coulomb forces and the relaxed-density ansatz for evaluation of proton and neutron densities of interacting nuclei. A simple prescription for the shell-correction part of the total potential is discussed. The parameters of the shell-correction and macroscopic parts of the relaxed-density potential are found by fitting the empirical barrier heights of the 89 nucleus-nucleus systems as well as  macroscopic potentials evaluated for 1485 nucleus-nucleus systems at 12 distances around touching points.
\end{abstract}


\maketitle

\section{Introduction}

Knowledge of the nucleus-nucleus potential around the barrier is necessary for the analysis of heavy-ion fusion and other nuclear reactions \cite{bass80,satchler,fl,dp,dn,dh,dts,dt,uo,sww}. Heavy-ion fusion at low energies is very important in the physics of stars \cite{dt,a,ddp}. As a result, the nucleus-nucleus interaction potential is a key ingredient for the description of various nuclear reactions in the nature.

The microscopic evaluation of a potential between nuclei is based on both the effective nucleon-nucleon force and the nucleon density distributions of interacting nuclei; see Refs. \cite{satchler,fl,dp,bbk,bs,m3y,sovgb,dn,dpb,ghdn,dnest1,uo,lw,kobo,jwls,uomr,wl,umar,dts,dt,simenel} and papers cited therein. There are also simple phenomenological parametrizations of the nucleus-nucleus potential \cite{bass80,fl,dp,bass73,cw,bass77,prox77,kns,aw,winther,prox2000,d2002,sww,duttp}, which are often used for the description of various heavy-ion reactions. The parameters of phenomenological potentials depend smoothly on the numbers of nucleons of interacting nuclei. When the nuclei approach each other, the energies of the nucleon single-particle levels of each nucleus are shifted and split due to the interaction of nucleons belonging to different nuclei. Therefore, the shell structures of both nuclei are changed at small distances between the nuclei.

The nucleon single-particle motion and shell effects are studied in the framework of detailed microscopic theories, which take into account the time-dependent transformation of the two-center field into a single-center one \cite{uo,uomr,wl,umar,dts,dt} and the dissipation of the collective motion. Such microscopic approaches are related to the complex numerical calculations. The goal of this study is to find a simple and easy-to-use approximation for the nucleus-nucleus  potential, which takes into account the bulk and shell effects.

The full energy of a nucleus consists of the sum of macroscopic and microscopic contributions according to the shell-correction method proposed by Strutinsky \cite{s}. The shell effect contribution has been successfully evaluated using the Strutinsky shell-correction prescription \cite{s,s72}, which is simpler than the full microscopic treatment. The shell-correction energy is an important ingredient of various modern nuclear physics models because the shell corrections strongly improve the modeling accuracy. Many key parameters such as atomic masses \cite{apdt,mnms}, fission process \cite{s,s72}, cluster decay of heavy nuclei \cite{msd,dclust}, and fusion cross section hindrance at deep sub-barrier energies \cite{dhindr} are well described in the frameworks of the macroscopic-microscopic models.

The contribution of the shell structure to the nucleus-nucleus potential has fully been ignored in phenomenological approaches. Therefore, it is desirable to find an improved phenomenological potential which takes into account the shell structure variation induced by the interaction between nucleons in the interacting nuclei. Such a potential should be more precise because it takes into account the unique shell-correction contribution of  interacting nuclei. The shell-correction energy varies from one nucleus to another substantially \cite{s}. In contrast, the macroscopic part of the full potential is described by a simple phenomenological expression which depends smoothly on the numbers of protons and neutrons of interacting nuclei. So, the macroscopic-microscopic potential should take into account both gross and individual properties of the specific nucleus-nucleus system.

A simple approach to the full nucleus-nucleus potential consisting of macroscopic and microscopic parts for a wide range of nucleus-nucleus systems is developed in the present paper. The parameters of the full potential are found using both data for empirical nucleus-nucleus barriers and the values of macroscopic nucleus-nucleus potentials around the barrier for various systems. The compilation of the data for empirical nucleus-nucleus barrier heights is given in Refs. \cite{duttp,was,nbdhgmh,mintjw,gg}. The macroscopic nucleus-nucleus potential is evaluated in the framework of the non-local extended Thomas-Fermi (ETF) energy-density functional theory and the relaxed-density approach for the description of the nucleon density distribution of two interacting nuclei around the touching distance. The energy-density functional takes into account the contributions of the nucleon kinetic energy and the effective nucleon-nucleon Skyrme and Coulomb forces \cite{bgh}. The ETF kinetic energy-density functional consists of the Thomas-Fermi term and all correction terms of the order $\hbar^4$ \cite{bgh}. Such an approach shows high accuracy for the description of atomic masses \cite{apdt,bgh}, nucleus-nucleus potentials for various systems \cite{dn,lw,jwls} and the cluster decay of heavy nuclei \cite{dclust}.

A detailed description of the macroscopic and shell-correction contributions to the full nucleus-nucleus potential is given in Sec. 2. The shape and parameters of the relaxed-density nucleus-nucleus potential with the shell-correction contribution are defined in Sec. 3. Sec. 4 is devoted to a discussion of the results and conclusions.

\section{Relaxed-density nucleus-nucleus potential}

In the shell-correction method \cite{s}, the full interaction potential energy of the system of interacting nuclei is written in the form (see also Refs. \cite{dclust,dhindr,dr96,iims})
\begin{eqnarray}
V_{\rm tot}(R) &=& V_{\rm macro}(R) + V_{\rm sh}(R)  \\ &=& [E_{12}(R) - E_1 - E_2 ] \nonumber \\ &+& [\delta E_{12}(R) - \delta E_1 - \delta E_2]. \nonumber
\end{eqnarray}
Here $E_1$, $E_2$, $ \delta E_1$ and $\delta E_2$ are the macroscopic and shell-correction energies of the non-interacting spherical nuclei 1 and 2, respectively. $E_{12}(R)$ and $\delta E_{12}(R)$ are the macroscopic and shell-correction energies of the interacting nuclei at distance $R$ between the mass centers of separated nuclei. The shell-correction energies $\delta E_{12}(R)$, $ \delta E_1$ and $\delta E_2$ include proton and neutron shell correction energies related to both the non-uniformity of single-particle spectra around the Fermi energies and the pairing corrections \cite{s}.

\subsection{Macroscopic interaction potential between nuclei}

The macroscopic part of the interaction potential energy $V_{\rm macro}(R)$ between spherical nuclei is given by
\begin{eqnarray}
V_{\rm macro}(R) = E_{12}(R) - E_1 - E_2 .
\end{eqnarray}
The macroscopic energies of nuclei in Eqs. (1) and (2) are evaluated with the help of a semiclassical ETF energy-density functional ${\cal E}[ \rho_{p}({\bf r}), \rho_{n}({\bf r})]$. These energies are
\begin{eqnarray}
E_{12}(R) = \int {\cal E}[\rho_{p}(\mathbf{r},R), \rho_{n}(\mathbf{r},R)] d{\bf r}, \\
E_k = \int \; {\cal E}[ \rho_{kp}({\bf r}), \rho_{kn}({\bf r})] \;
d{\bf r},
\end{eqnarray}
where $\rho_{1p}({\bf r})$, $\rho_{2p}({\bf r})$, $\rho_{1n}({\bf r})$ and $\rho_{2n}({\bf r})$ are the proton and neutron densities of the non-interacting nuclei, while $\rho_{p}(\mathbf{r},R)$ and $\rho_{n}(\mathbf{r},R)$ are the proton and neutron densities of the interacting nuclei, $k=1,2$.

\subsubsection{Energy-density functional}

The semiclassical energy-density functional includes the Skyrme and Cou\-lomb interactions as well as the intrinsic proton and neutron kinetic energies obtained in the ETF approach \cite{bgh}. According to \cite{bgh} the following expression for the energy density functional has been deduced:
\begin{eqnarray}
{\cal E}[ \rho_{p}({\bf r}), \rho_{n}({\bf r})] =
\frac{\hbar^2}{2m} [\tau_p({\bf r}) +\tau_n({\bf r})] + {\cal
V}_{\rm Sk}({\bf r}) + {\cal V}_{\rm C}({\bf r}).
\end{eqnarray}
The kinetic parts for protons ($i=p$) and
neutrons ($i=n$) are given by
\begin{eqnarray}
\tau_{i}({\bf r}) = \tau_{i TF}({\bf r}) + \tau_{i 2}({\bf r}) + \tau_{i 4}({\bf r}),
\end{eqnarray}
where $\tau_{i TF}({\bf r})$ is the Thomas-Fermi contribution to the kinetic energy-density functional, and $\tau_{i 2}({\bf r})$ and $\tau_{i 4}({\bf r})$ are semiclassical $\hbar^2$ and $\hbar^4$ correction terms to the kinetic energy-density functional for the non-local case, respectively.

The nuclear interaction part ${\cal V}_{\rm Sk}({\bf r})$ results from the Skyrme force and reads
\begin{eqnarray}
{\cal V}_{\rm Sk}({\bf r}) \; = \; \frac{t_0}{2} \;
[(1+\frac{1}{2}x_0) \rho^2 -
(x_0+\frac{1}{2}) (\rho_p^2+\rho_n^2)] \;\\
+\frac{1}{12} t_3 \rho^\alpha [(1+\frac{1}{2}x_3 )\rho^2 -
(x_3+\frac{1}{2}) (\rho_p^2+\rho_n^2) ] \nonumber \\
+\frac{1}{4} [t_1(1+\frac{1}{2}x_1)+t_2(1+\frac{1}{2}x_2)] \tau \rho
\nonumber \\
+\frac{1}{4} [t_2(x_2+\frac{1}{2}) - t_1(x_1+\frac{1}{2})] (\tau_p
\rho_p+\tau_n \rho_n)
\nonumber \\
+\frac{1}{16}[3t_1(1+\frac{1}{2} x_1)-t_2(1+\frac{1}{2}x_2)] (\nabla
\rho)^2 \nonumber \\
- \frac{1}{16}[3t_1(x_1+\frac{1}{2} )+t_2(x_2+\frac{1}{2})] ((\nabla
\rho_n)^2 +(\nabla \rho_p)^2 ) \nonumber \\
- \frac{W_0}{2} \left[ \mathbf{J} \cdot \mathbf{\nabla}\rho
+ \mathbf{J}_n \cdot \mathbf{\nabla}\rho_n
+ \mathbf{J}_p \cdot \mathbf{\nabla}\rho_p \right] , \nonumber
\end{eqnarray}
where $t_0$, $t_1$, $t_2$, $x_0$, $x_1$, $x_2$, $\alpha$ and $W_0$
are the Skyrme-force parameters, $\mathbf{J}_i$ are the spin-orbit densities, $\rho=\rho_p+\rho_n$, $\tau=\tau_p+\tau_n$ and $\mathbf{J}= \mathbf{J}_p + \mathbf{J}_n$. The expressions for $\tau_{i 2}({\bf r})$, $\tau_{i 4}({\bf r})$, and $\mathbf{J}_i$ for a non-local case are given in Ref. \cite{bgh}.

The Coulomb-energy density is given by
\begin{eqnarray}
{\cal V}_{\rm C}({\bf r}) = \frac{e^2 }{2} \rho_p({\bf r}) \int \;
\frac{\rho_p ({\bf r}' )}{|{\bf r}-{\bf r}' |}
d {\bf r}' \nonumber \\
-\frac{3e^2}{4} \left( \frac{3}{\pi} \right)^{1/3} [\rho_p({\bf
r})]^{4/3},
\end{eqnarray}
where the first term is the direct contribution, the second term is the local approximation to the exchange contribution, and $e$ is the proton charge.

This approximation for the energy of nucleus based on the semiclassical energy-density functional is simple and accurate. Similar approximations ware successfully used for the evaluation of the atomic masses \cite{apdt}, the fission barrier characteristics \cite{bgh}, the nucleus-nucleus potentials \cite{dn,dpb,dnest1,lw,d2002,d_barr} and the cluster emission \cite{dclust}.

\subsubsection{Parametrization of density distribution}

The nucleon density is a necessary ingredient of the semiclassical energy-density approach; see Eqs. (3)--(8). It is difficult to find the proton and neutron density distributions by solving the integro-differential variational Lagrange equations in the framework of the non-local $\hbar^2$ ETF approach for the case of a spherical nucleus \cite{dnest2}. The neutron and proton densities of nuclei in the case of the non-local $\hbar^4$ ETF approach are found by using trial functions only \cite{bgh}. The proton and neutron density distributions for system of interacting nuclei have not been evaluated  yet in the framework of the ETF.

As a rule the densities of interacting nuclei are parametrized according to specific physical conditions of the reaction. The frozen approximation for the density distributions of interacting nuclei is often used for the evaluation of the nucleus-nucleus potentials in the frameworks of the energy-density \cite{bbk,dn,dpb,dnest1,lw,jwls} and double-folding \cite{satchler,fl,m3y,sovgb,kobo} approaches. Note that it is reasonable to use the frozen approximation for fast nucleus-nucleus collisions, when the proton and neutron densities cannot significantly relax during the collision process \cite{dn}. The proton or neutron densities at the fixed point of space in this approximation are the sum of the corresponding nucleon densities of each nucleus in this point. As the result, the nucleon density can exceed the equilibrium density of the nuclear matter $\overline{\rho}$ in some space region at small distances between nuclei.

The nucleus-nucleus fusion reactions around the barrier and, especially, deeply below it are related to the slow process of collective motion of nuclei. During this process the proton and neutron densities can relax and the densities of nuclei cannot be simply presented as sums of nucleon densities of the interacting nuclei. The relaxed-density distributions should fulfill obvious conditions:
\begin{itemize}
\item[--] The values of density at any point in space do not exceed the equilibrium density of nuclear matter $\overline{\rho}$ because the compressibility modulus of nuclear matter strongly prevents excess of $\overline{\rho}$.
\item[--] The values of relaxed density at any point in space should be smaller than the one in the frozen approximation but larger than the density values of any of the interacting nuclei at this point.
\end{itemize}
Taking into account these conditions, the relaxed-density ansatz \cite{dclust} for the proton (neutron) density of the interacting nuclei at point $\mathbf{r}$ can be presented as
\begin{eqnarray}
\rho_{p(n)}(\mathbf{r},R) &=&
\rho_{1p(n)}(\mathbf{r})
+\rho_{2p(n)}(\mathbf{r},R)  \\ \nonumber
&-& \frac{2 \; \rho_{1p(n)}(\mathbf{r}) \;
\rho_{2p(n)}(\mathbf{r},R)}
{\rho_{1p(n)} +\rho_{2p(n)}}.
\end{eqnarray}
Here $\rho_{1p(n)}(\mathbf{r})=\rho_{1p(n)} f_{1p(n)}(\mathbf{r})$ and $\rho_{2p(n)}(\mathbf{r},R)=\rho_{2p(n)} f_{2p(n)}(\mathbf{r},R)$ are the density distributions of the proton (neutron) subsystem related to the first and second nuclei, respectively. The functions $f_{1p(n)}(\mathbf{r})$ and $f_{2p(n)}(\mathbf{r},R)$ describe the space distributions of corresponding densities. These functions are normalized to unity at the center of corresponding nuclei $f_{1p(n)}(\mathbf{r})|_{r=0}=f_{2p(n)}(\mathbf{r},R)|_{r=0}= 1$.

Let us consider the cases of strongly and weakly overlapping nuclei in the framework of  parametrization (9). For the sake of simplicity, similar shapes of the proton and neutron density distributions in each nucleus are proposed for consideration of these cases. Therefore, the density distributions are
\begin{eqnarray*}
\rho_{1p}(\mathbf{r})=\frac{Z_1}{A_1}\overline{\rho}  f_{1}(\mathbf{r}), \;\; \rho_{1n}(\mathbf{r})=\frac{N_1}{A_1}\overline{\rho} f_{1}(\mathbf{r}), \\  \rho_{2p}(\mathbf{r},R)=\frac{Z_2}{A_2}\overline{\rho} f_{2}(\mathbf{r},R), \;\; \rho_{2n}(\mathbf{r},R)=\frac{N_2}{A_2}\overline{\rho} f_{2}(\mathbf{r},R),
\end{eqnarray*}
where $Z_i$ and $N_i$ are the numbers of proton and neutron in the nucleus $i$, $A_i=Z_i+N_i$, $i=1,2$.

The nuclei overlap strongly at small distances between them. The total densities of each nucleus at the point close to the center of each nucleus are
\begin{eqnarray*} \rho_{1}(|\mathbf{r}|)= \overline{\rho}(1-g_1), \\ \rho_{2}(|\mathbf{r}|,R)=\rho_{2}(R-|\mathbf{r}|)= \overline{\rho}(1-g_2),
\end{eqnarray*}
where $g_{1}=1-f_{1}(|\mathbf{r}|) \ll 1$ and $g_{2}=1- f_{2}(|\mathbf{r}|,R) \ll 1$.
Substituting these densities into (9) I find the value of relaxed density at small distance between nuclei at the point between the centers:
\begin{eqnarray*}
\rho(\mathbf{r},R) \approx \overline{\rho}[1-(g_1+g_2)/2] < \overline{\rho}.
\end{eqnarray*}
In contrast to this, the total density of these nuclei at the same point in the frozen density approach is
\begin{eqnarray*}
\rho_{\rm frozen}(\mathbf{r},R) = \rho_{1}(|\mathbf{r}|)+ \rho_{2}(|\mathbf{r}|,R) = \overline{\rho}[2-g_1-g_2] > \overline{\rho}.
\end{eqnarray*}
Note that
\begin{eqnarray*}
\rho(\mathbf{r},R) \approx \overline{\rho}[1-(g_1+g_2)/2] < \rho_{\rm frozen}(\mathbf{r},R) .
\end{eqnarray*}

In the opposite case of well separated nuclei (large distance between interacting nuclei) the nucleon densities of each nucleus in the point between nuclei and far from the center of each nucleus are very small
\begin{eqnarray*}
\rho_{1}(|\mathbf{r}|) = \overline{\rho}  f_{1}(|\mathbf{r}|) =\overline{\rho}  f_{1}^0,
\end{eqnarray*}
\begin{eqnarray*}
\rho_{2}(|\mathbf{r}|,R) =  \overline{\rho} f_{2}(|\mathbf{r}|,R) = \overline{\rho} f_{2}^0,
\end{eqnarray*}
where $f_{1(2)}^0 \ll 1$. Substituting these densities into (9) I obtain the value of relaxed density at this point $$\rho(\mathbf{r},R) \approx \overline{\rho}(f_{1}^0+f_{2}^0 - f_{1}^0 f_{2}^0).$$ Comparing this density with the other ones, it is easy to conclude that
\begin{itemize}
\item[--] The value of relaxed density is larger than the density of each nucleus at this point, i.e.
\begin{eqnarray*}
    \rho(\mathbf{r},R) > \rho_{1}(|\mathbf{r}|) = \overline{\rho} f_{1}^0, \;\;  \rho(\mathbf{r},R) > \rho_{2}(|\mathbf{r}|,R) = \overline{\rho} f_{2}^0.
\end{eqnarray*}

\item[--] The value of relaxed density is smaller than the sum of densities of these nuclei at this point, i.e.
\begin{eqnarray*}
    \rho(\mathbf{r},R) \approx \overline{\rho}(f_{1}^0+f_{2}^0 - f_{1}^0 f_{2}^0) \leq \rho_{1}(\mathbf{r}) + \rho_{2}(\mathbf{r},R) \\ = \overline{\rho}(f_{1}^0+f_{2}^0) = \rho_{\rm frozen}(\mathbf{r},R).
\end{eqnarray*}
\end{itemize}

Thus, the parametrization (9) satisfies the proposed conditions. This para\-metrization leads to a more realistic spatial distribution of density formed by two nuclei near the touching point at collision energies around the barrier height than the one obtained in the frozen density approach. Approximation (9) also drastically simplifies the numerical calculations of the macroscopic potential $V_{\rm m}(R) $ in the framework of the ETF approach with the Skyrme and Coulomb forces.

The proton (neutron) density distributions of separated spherical nuclei are evaluated in the framework of the Hartree-Fock method using the SLy4 parameter set of the Skyrme force \cite{sly4}. The BCS pairing with the Lipkin-Nogami method \cite{ln} for approximate particle number projection is used in the calculation. The microscopic proton and neutron densities are given on the mesh with step $h$. At small distances between nuclei, when the densities of nuclei are overlapped, let us propose that the proton and neutron density values of each nucleus are the same as before at any mesh points. However, the size of the mesh is slightly changed to $\alpha_{p(n)} h$. The scaling factors $\alpha_{p(n)}$ can be fixed at any value of $R$ by the conservation conditions of the total numbers of protons $Z_1+Z_2$ and neutrons $N_1+N_2$ in the system of interacting nuclei, which are
\begin{eqnarray}
\int d \mathbf{r}\; \rho_{p(n)}(\mathbf{r},R) = Z_1+Z_2(=N_1+N_2).
\end{eqnarray}
The values of $\alpha_{p(n)}$ are very close to 1 at $R$ around the touching point of the nuclei.

By using ansatz (9) for the relaxed-density parametrization, the microscopic Hartree-Fock densities and conditions (10), and the description of densities of two nuclei at large $R$ and around the touching point is simplified. As a result, the macroscopic interaction potential energy of the nuclei $V_{\rm macro}(R)$ can be easily obtained by substituting these densities into Eqs. (2)--(8).

\subsection{Shell-correction contribution to the full interaction potential between nuclei}

The shell-correction contribution to the full interaction potential energy is
\begin{eqnarray}
V_{\rm sh}(R) = \delta E_{12}(R) - \delta E_1 - \delta E_2.
\end{eqnarray}

It is obvious that mutual influence of nuclei on their single-particle spectra is negligible at distances much greater than the sum of radii of nuclei $R_t=R_1+R_2$, therefore
\begin{eqnarray}
\delta E_{12}(R)|_{R \gg R_t} = \delta E_1 + \delta E_2
\end{eqnarray}
and
\begin{eqnarray*}
V_{\rm sh}(R) |_{R \gg R_t} = 0 .
\end{eqnarray*}

The $\delta E_{12}(R)$ value at $R \approx 0$ equals the shell-correction energy of the nucleus formed in the fusion of nuclei 1 and 2 at the equilibrium deformation. The full potential should be equal to the fusion reaction $Q$-value at $R=0$, {\it i.e.} $V_{\rm tot}(R=0)=Q$.

The nuclei strongly interact at small distances between them. This interaction leads to the shift and the splitting of the single-particle levels in both nuclei. Due to this the proton and neutron single-particle spectra in the vicinity of the respective Fermi levels become more homogenous around touching distance $R_{t}$. Such behavior of the single-particle levels is clearly demonstrated in the frameworks of the two-center shell model \cite{dts,dt,add,mg} and microscopic cluster decay model \cite{msd}.

According to the shell-correction method \cite{s,s72}, the absolute value of the shell correction energy is reduced in the case of the more homogeneous single-particle spectrum in the vicinity of the Fermi level. The sharp reduction of the shell-correction contribution to the full potential energy around the touching point of nuclei is obtained in the framework of the microscopic cluster decay model \cite{msd}. So, the perturbation of the single-particle nucleon levels, which is proportional to the strength of nucleon-nucleus interaction, reduces the absolute value of the shell-correction energy.

The perturbation of the single-particle level energies and the interaction between nucleons belonging to different nuclei are increased when the distance between the surfaces of the interacting nuclei decreases. The perturbation strength is related to the density distribution in the nucleus induced by the disturbance as well as to the radius of the nucleon-nucleon force. The density distribution is often parametrized by the Fermi distribution. Since the radius of the nucleon-nucleon force is very short, it is reasonable to approximate the shell-correction contribution to the full nucleus-nucleus potential around the touching point of two nuclei as \cite{dclust,dhindr}
\begin{eqnarray}
\delta E_{12}(R) \approx [\delta E_1 + \delta E_2 ] f_{\rm sh}(R).
\end{eqnarray}
Here
\begin{eqnarray}
f_{\rm sh}(R) = 1/\left\{1 + \exp{[(R_{\rm sh}-R)/d_{\rm sh}}]\right\},
\end{eqnarray}
and $R_{\rm sh}$ and $d_{\rm sh}$ are the radius and the diffuseness related to the attenuation of the shell-correction magnitude with reduction of distance $R$. This approximation for the shell-correction energy contribution into the full potential is rough, but it can greatly simplify the calculations of the shell-correction energies for various nucleus-nucleus systems at different distances $R$. Note that numerical calculations of the shell-correction energy in the framework of the two-center model \cite{msd} show that the values of the shell-correction energy of the nucleus-nucleus system abruptly change around the touching point of two nuclei. Therefore, the radial dependence of the shell-correction energy described by Eqs. (13) and (14) is reasonable. Moreover, the exponential reduction of the shell-correction energy values related to washing out the non-homogeneity of the single-particle spectra is often considered in nuclear physics \cite{dp,ms,ist,dh}.

The shell-correction energy of interacting nuclei $\delta E_{12}(R)$ smoothly approaches the limit of non-interacting nuclei (12) at large distances $R$ between nuclei, {\it i.e.}, $V_{\rm sh}(R) \rightarrow 0$ at $R \gg R_{\rm sh} \sim R_t$. As a result, the shell-correction contribution to the full nucleus-nucleus potential is
\begin{eqnarray}
V_{\rm sh}(R) \simeq [\delta E_1 + \delta E_2 ] \cdot [f_{\rm sh}(R)-1].
\end{eqnarray}
This approximation for the shell-correction contribution to the potential energy of two nuclei is successfully used for the description of both the cluster emission from heavy nuclei \cite{dclust} and the fusion cross section hindrance at deep sub-barrier energies \cite{dhindr}.

It is possible to evaluate the shell-correction energy values $\delta E_1$ and $\delta E_2$ according to the Strutinsky prescription \cite{s,s72} for a specific nucleon mean-field. However, the easiest way to estimate the value of the shell-correction energy $\delta E$ in a spherical nucleus is to find the difference between the experimental $B_{\rm exp}$ and macroscopic $B_{\rm m}$ binding energies of the nucleus:
\begin{eqnarray}
\delta E = B_{\rm exp} - B_{\rm m}.
\end{eqnarray}
Moreover, this approach for the evaluation of the shell-correction energy is similar to the one applied for the evaluation of the energy level density in nuclei; see Refs. \cite{ripl3,mn}.

The values of $B_{\rm exp}$ in Eq. (16) can be found in the recent evaluation of the atomic masses \cite{ame2012}, while the value of $B_{\rm m}$ in the nucleus with $Z$ protons and $N$ neutrons is
\begin{eqnarray}
B_{\rm m} = -15.86864 A+21.18164 A^{2/3}-6.49923 A^{1/3}  + \nonumber \\
 \left[\frac{N-Z}{A}\right]^2 \left[26.37269 A -23.80118 A^{2/3} - 8.62322 A^{1/3} \right] \nonumber \\
 + \frac{Z^2}{A^{1/3}} \left[ 0.78068- 0.63678 A^{-1/3} \right] + P_p+P_n. \;\;\;
\end{eqnarray}
Here $B_{\rm m}$ is the binding energy in MeV, $A=Z+N$, $P_{p(n)}$ are the proton (neutron) pairing terms, which equal to $P_{p(n)}=5.62922 (4.99342) A^{-1/3}$ in the case of odd $Z$ ($N$) and $P_{p(n)}=0$ in the case of even $Z$ ($N$). Equation (17) is a simple extension of the liquid drop Weizs\"acker formula \cite{weiz,bm} for the binding energy of nuclei. The coefficients in Eq. (17) are evaluated by fitting the recent values of the atomic masses \cite{ame2012}. The experimental binding energies of 3353 nuclei are described by Eq. (17) with a root mean error of 2.49 MeV. This error is very small compared to the experimental values of the nuclear binding energies in medium and heavy nuclei.

The liquid drop mass formula may be found by using the Skyrme force too \cite{rbnw}. Such an approach may look more consistent because the macroscopic nucleus-nucleus interaction energy is evaluated with the help of the Skyrme energy density functional; see Eqs. (2)-(8).  However, the liquid drop mass formula obtained in Ref. \cite{rbnw} using the Hartree-Fock calculation with the SLy4 parametrization of the Skyrme force leads to a larger value of the root mean error for experimental masses than the one in Eq. (17). The atomic masses of very heavy nuclei calculated in the framework of the Hartree-Fock approach with the SLy4 parameter set are systematically underbound \cite{dsn}. Therefore, Eqs. (16) and (17) are applied for the evaluation of the shell-correction energy here because it is better to use a simple and high accuracy approach.

Note that slightly different values of the shell-correction energy for the same nucleus can be calculated applying various formulas for the liquid drop mass formula \cite{mn} or using diverse parametrizations of the nucleon mean field for the shell-correction calculations according to the Strutinsky prescription. Similarly, the Hartree-Fock calculations with various parameter sets of the Skyrme force lead to different values of the shell-correction energy for the same nucleus. For example, the values of the shell-correction energy of $^{208}$Pb evaluated by using Eqs. (16)-(17), the Megnoni-Nakamura approach \cite{mn}, and the Nilsson and Saxon-Woods mean-field models \cite{s72,sclv} are $-10.620$, $-9.972$, $-12$ and $-12$ MeV, respectively. These different values of the shell-correction energy are slightly affected by the value of the nucleus-nucleus potential due to the radial dependence of the shell-correction contribution, see Eqs. (14) and (15). For example, the variations of the nucleus-nucleus potentials for the system $^{16}$O+$^{208}$Pb at the touching point induced by various values of the shell-correction energies of $^{208}$Pb ($-10.620$ and $-9.972$ MeV or $-10.620$ and $12$ MeV) are 0.16 or 0.34 MeV, respectively. Moreover, the corresponding differences of the nucleus-nucleus potentials near the barrier distances are even smaller than 0.16 or 0.34 MeV due to the exponential dependence $f_{\rm sh}(R)$ on $R$. Note that the value of the barrier height of the nucleus-nucleus potential for the system $^{16}$O+$^{208}$Pb is close to 77 MeV \cite{dhindr}. Therefore, the model-dependent variation of the shell-correction energy is not crucial for the accuracy of the nucleus-nucleus potential.

Any influence of the relative motion of the nuclei on the nucleus-nucleus interaction and the single-particle levels has been ignored up to now. However, the nucleons move in the approaching nuclei during reaction and the interaction of nucleons belonging to different nuclei disturbs the nucleons in both nuclei. This leads to time-dependent disturbance of the shell structure of each nucleus. It is obvious that the strength of this disturbance should depend on the ratio of the nucleon velocity (kinetic energy) and the relative velocity (kinetic energy) of the nuclei. The relative velocity of two nuclei depends on radial and rotational degrees of freedom. When the relative velocity of approaching nuclei is small, the static (adiabatic) consideration is close to realistic because the shell structure of the nuclei can be relaxed due to high velocities of the nucleons in the nuclei. In contrast, the shell structure and nucleon density distributions cannot be disturbed at fast collisions. Therefore, nuclei can touch each other without modification of the shell structure of colliding nuclei at high collision energies. This conclusion is supported by time-dependent microscopic calculations \cite{wl,umar}, which show that the nuclei overcome the barrier essentially in their ground-state density at high collision energies. Thus, the shell-correction contribution to the full potential depends on the collision energy $E$. The energy dependence of this contribution can be presented in the form \cite{dhindr}
\begin{eqnarray}
V_{\rm sh}(R,E) = \left\{ \begin{array}{lr} V_{\rm sh}^0(R) \cdot \exp{[-a(E-B)]}, & {\rm at} \; E \geq B, \\
V_{\rm sh}^0(R), & {\rm at} \; E \leq B, \\
\end{array} \right.
\end{eqnarray}
where $B$ is the barrier height of the macroscopic potential $V_{\rm macro}(R)$ and $a$ is the reduction parameter of the shell-correction energy contribution to the full potential. So, the shell-correction contribution to the full potential decreases with increasing the collision energy. Eq. (18) does not depend on the velocity of the nucleons because nucleon velocity at the Fermi level has a small variation from one nucleus to another and the value of the shell-correction energy depends on the inhomogeneity of the single-particle spectra around the Fermi energy \cite{s}.

The full potential at $E>B$ moves smoothly towards the macroscopic potential with increasing $E$. The nucleus-nucleus collision process is near adiabatical at small collision energies $E \lesssim B$, because the velocity of relative motion is less than the velocities of nucleons in nuclei. Therefore, the energy dependence of the shell-correction contribution is ignored at low collision energies.

The value of the parameter $a=0.35$ MeV$^{-1}$ in Eq. (18) is determined by fitting the fusion cross section for the reaction $^{16}$O + $^{208}$Pb \cite{dhindr} because it has been measured in a very wide energy range around the barrier \cite{OPb}. The thicknesses of the barrier at deep sub-barrier collision energies become larger due to the positive shell-correction contribution to the full potential for the reaction $^{16}$O + $^{208}$Pb. This directly leads to a reduction of the barrier penetrability and, as a result, to the hindrance of the deep sub-barrier fusion which is observed for this reaction \cite{dhindr,OPb}.

\section{Phenomenological relaxed-density nucleus-nucleus potential}

The proton and neutron densities of the nuclei are obtained in the framework of the Hartree-Fock model and the relaxed-density ansatz (9). The macroscopic relaxed-density nucleus-nucleus potential is calculated substituting these densities into Eqs. (2)-(8). The parameter set SLy4 \cite{sly4} of the Skyrme force is used for the evaluation of both the Hartree-Fock nucleon densities and the macroscopic part of the nucleus-nucleus potential. This parameter set is adopted in the calculations because the nuclei along and far away from the beta-stability line are often used in heavy-ion reactions and this set of the Skyrme force is adjusted to reproduce various properties of nuclear matter, neutron stars, and finite nuclei along and far away from the beta-stability line.

The values of fusion barrier heights obtained in the macroscopic relaxed-density approach for light and medium nucleus-nucleus systems are close to the empirical ones from Refs. \cite{duttp,nbdhgmh,gg}. Unfortunately, the evaluated values of the fusion barrier heights for heavy and very heavy nucleus-nucleus systems are higher than the empirical ones. Note that a similar tendency is found in Ref. \cite{nbdhgmh} when the barrier heights of nucleus-nucleus potentials are obtained in the framework of the semiclassical energy-density approach with the frozen density distribution of two nuclei \cite{d2002} and the SkM$^\star$ parameter set of the Skyrme force \cite{skmstar} (see also Fig. 1).  Such a tendency is probably related to a systematical underbounding \cite{dsn} of very heavy nuclear systems and/or nuclei in the framework of the Skyrme-Hartree-Fock approach. Moreover, the overestimation of empirical barrier heights for heavy and very heavy nucleus-nucleus systems is observed for various phenomenological potentials too; see Fig. 1.

Our goal is to find the phenomenological relaxed-density potential which describes well the empirical barriers for light, medium, heavy, and very heavy nucleus-nucleus systems. The values of the macroscopic part of this potential around the touching distances should be close to the numerically evaluated relaxed-density macroscopic potential. To solve this task the next approach is applied.

At the beginning, 54 spherical nuclei between $^{16}$O and $^{216}$Po are selected. Then 1485 macroscopic relaxed-density potentials between all possible pairs of these 54 nuclei at 12 distances around the touching point are evaluated. These distances equal $R_{i,j}=1.18( A_{1i}^{1/3}+A_{2i}^{1/3})+0.3 j-0.6$ fm, where $A_{1i}$ and $A_{2i}$ are the numbers of nucleons in the interacting nuclei, $i=1,2,3,..., 1485$ and $j=1,2,3,..., 12$. The dataset for 17820 ($17820=1485\times12$) macroscopic potential values $V_{i{\rm macro}}(R_{ij})$ and the data for 89 empirical barrier heights $B_k^{\rm emp}$ between spherical nucleus-nucleus systems from Refs. \cite{duttp,nbdhgmh,mintjw,gg} are used for the evaluation of the macroscopical and shell-correction parts of the full phenomenological relaxed-density nucleus-nucleus potential. Note that, as a rule, the values of the empirical barrier heights are extracted from the analysis of subbarrier heavy-ion fusion reactions. Therefore, the shell-correction contributions into the full nucleus-nucleus potentials are evaluated in the case $E\leq B$. The parameters of the phenomenological potential are found by the minimization of the function
\begin{eqnarray}
F &=& 15 \left[ \frac{1}{17820} \sum_{i=1}^{1485} \sum_{j=1}^{12}
\left( \frac{V_{i{\rm macro}}^{\rm phen}(R_{ij})}{V_{\rm macro}^i(R_{ij})} - 1 \right)^2 \right]^{1/2} \nonumber \\ & & + \left[ \frac{1}{89}\sum_{k=1}^{89} \left( B_k^{\rm emp}-B_k^{\rm phen} \right)^2\right]^{1/2}.
\end{eqnarray}
Here $B_k^{\rm phen}$ are the values of the barrier heights evaluated by using the full phenomenological potential. The values of the macroscopical part of the phenomenological potential should be similar to the ones of the numerically evaluated relaxed-density potential due to the minimization of the first term in Eq. (19). The minimization of the second term of Eq. (19) leads to correct values of barriers for various nucleus-nucleus systems. The factor 15 in the first term of Eq. (19) enhances the influence of fitting of the macroscopic potentials, and the values of the first and second terms of Eq. (19) are close to each other in this case.

The full phenomenological relaxed-density nucleus-nucleus potential is the sum of the Coulomb and nuclear parts
\begin{eqnarray}
V_{\rm tot}(R) = V_{\rm C}(R) + V_{\rm nucl}(R),
\end{eqnarray}
where
\begin{eqnarray}
V_{\rm C}(R) = \left\{ \begin{array}{lr}
\frac{Z_1 Z_2 e^2}{R}, & {\rm at} \; R \geq R_{\rm C}, \\
\frac{Z_1 Z_2 e^2}{R_{\rm C}} \left[ \frac{3}{2} -\frac{1}{2} \; \frac{R^2}{R_{\rm C}^2} \right], & {\rm at} \; 0 \leq R \leq R_{\rm C} , \\
\end{array} \right.
\end{eqnarray}
\begin{eqnarray}
V_{\rm nucl}(R) = \left\{ \begin{array}{lr}
V_{\rm macro}^{\rm out}(R) + V_{\rm sh}(R), & {\rm at} \; R \geq R_{\rm t}, \\
V_{\rm nucl}^{\rm in}(R), & {\rm at} \; 0 \leq R \leq R_{\rm t}, \\
\end{array} \right.
\end{eqnarray}
and
\begin{eqnarray}
R_i=1.2536 A_i^{1/3}-0.80012 A_i^{-1/3}  -0.0021444/A_i
\end{eqnarray}
are the radii of corresponding nuclei in fm and $i=1,2$. The expression for the Coulomb potential is standard for the theory of heavy-ion reactions \cite{bass80,satchler,fl,dp}. Let us put $R_{\rm C}=R_{\rm t}=R_1+R_2$ for the sake of the reduction of the parameter number.

The macroscopic nuclear part $V_{\rm macro}^{\rm out}(R)$ of the potential is taken in the Woods-Saxon form
\begin{eqnarray}
V_{\rm macro}^{\rm out}(R) = \frac{V_0}{1+\exp[(R-R_{\rm t})/d]} ,
\end{eqnarray}
where
\begin{eqnarray}
V_0= v_1 C+ v_2 C^{1/2}
\end{eqnarray}
is the potential strength in MeV, $v_1=-27.190$ MeV fm$^{-1}$, $v_1=-0.93009$ MeV fm$^{-1/2}$,
\begin{eqnarray}
d=d_1 + d_2/C
\end{eqnarray}
is the diffuseness in fm, $d_1=0.78122$ fm, $d_2= - 0.20535$ fm$^2$ and $C=R_1 R_2/R_{\rm t}$ is in fm.

The shell-correction contribution $V_{\rm sh}(R)$ to the total potential at $R \geq R_t$ is evaluated using Eqs. (14)--(18). The shell-correction contribution to the potential is small near the barrier radius and important close to the touching point. Therefore, it is impossible to fix the parameters of the shell-correction potential unambiguously by using our procedure. However, the values of the deep sub-barrier fusion cross sections are sensitive to the shell-correction contribution of the heavy-ion potential \cite{dhindr}. Therefore, the radius $R_{\rm sh}$ and diffuseness $d_{\rm sh}$ parameters of the shell-correction contribution [see Eq. (14)] can be determined using the data for the  fusion cross section hindrance at deep sub-barrier energies for three reactions discussed in Ref. \cite{dhindr}. The values of these parameters are located in narrow ranges for various fusion reactions \cite{dhindr}. So, I can use the averaging of the corresponding values obtained in Ref. \cite{dhindr} for fixing values of $R_{\rm sh}$ and $d_{\rm sh}$. The average value of the diffuseness parameter is $d_{\rm sh} = 0.233$ fm. Parameter $R_{\rm sh}$ can be related to the touching distance of nuclei $R_{\rm t}$, {\it i.e.}, $R_{\rm sh}= R_{\rm t} - 0.26$ fm.

The nuclear part of the phenomenological potential at small distances is also taken in the Woods-Saxon form
\begin{eqnarray}
V_{\rm nucl}^{\rm in}(R)=\frac{Q_{\rm eff}}{1+\exp[(R-R_{\rm in})/d_{\rm in}]} ,
\end{eqnarray}
where
\begin{eqnarray}
Q_{\rm eff} = Q - \frac{3Z_1 Z_2 e^2}{2R_{\rm t}}, \\
R_{\rm in}= R_{\rm t}-d_{\rm in} \log\left(\frac{Q_{\rm eff}}{V_{\rm nucl}(R_{\rm t})}-1 \right) , \\
d_{\rm in} =-\frac{V_{\rm nucl}^2(R_{\rm t})}{Q_{\rm eff} V_{\rm nucl}^\prime(R_{\rm t})} \left(\frac{Q_{\rm eff}}{V_{\rm nucl}(R_{\rm t})}-1 \right) .
\end{eqnarray}
Here $Q$ is the $Q$-value of the heavy-ion fusion reaction evaluated by using the recent values of the atomic masses \cite{ame2012}. If the binding energy of a nucleus is not given in Ref. \cite{ame2012}, then it can be obtained with the help of Eq. (17). The radius $R_{\rm in}$ and the diffuseness $d_{\rm in}$ of the inner nuclear potential are obtained by using the continuity conditions of both the potential
\begin{eqnarray}
 V_{\rm nucl}(R_{\rm t})=V_{\rm macro}^{\rm out}(R_{\rm t}) + V_{\rm sh}(R_{\rm t})= V_{\rm nucl}^{\rm in}(R_{\rm t})
\end{eqnarray}
and the potential derivative
\begin{eqnarray}
V_{\rm nucl}^\prime (R_{\rm t})= {V_{\rm nucl}^{\rm in}}^\prime (R_{\rm t})
\end{eqnarray}
at the matching point $R=R_{\rm t}$.

The nucleus-nucleus interaction at large distances $R \geq R_{\rm t}$ contains the macroscopic nuclear, Coulomb, and shell-correction contributions, see Eqs. (20)--(22). In the general case, the heavy-ion potential energy at distances much smaller than $R_{\rm t}$ depends on the shape of fusing nuclei. For the sake of simplicity, the potentials in Eqs. (24) and (27) are parametrized by the Woods-Saxon one, because it is very often used in the theory of nuclear reactions. Therefore, this potential can be easily integrated into existing nuclear reaction codes.

According to Eq. (1) the potential energy of a nucleus formed at complete fusion of two nuclei equals the reaction $Q$-value, {\it i.e.}, $V_{\rm tot}(0) = Q$. The relaxed density potential is approximately equal to $Q$ at $R=0$ due to Eqs. (20)--(22), and (27)--(30).

\section{Results and discussion}

\begin{figure}
\begin{center}
\includegraphics[width=8.7cm]{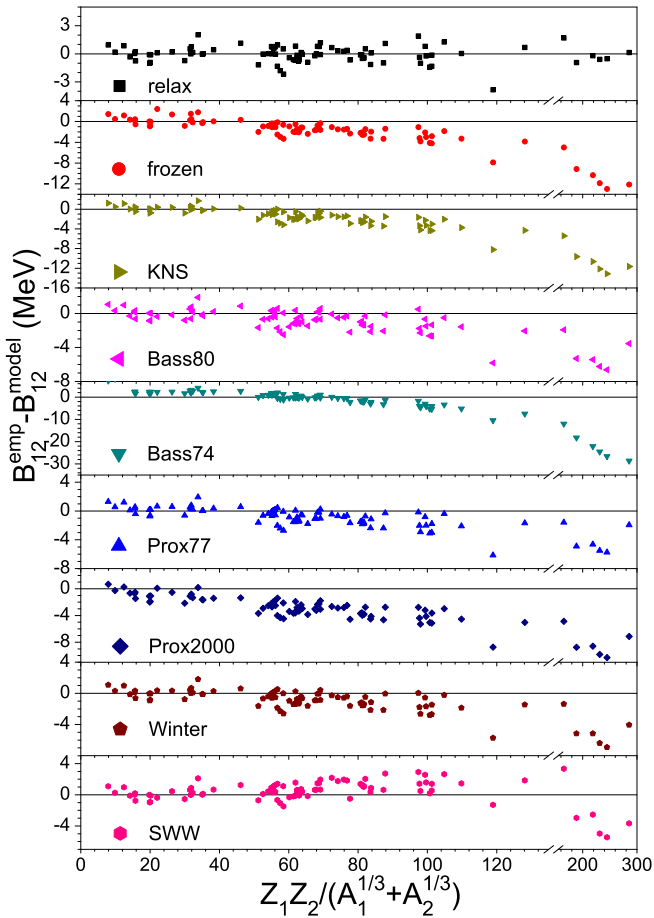}
\caption{"(Color online)
Absolute differences between 89 empirical barrier heights \cite{duttp,nbdhgmh,mintjw,gg} and barrier heights evaluated using the relaxed density with the shell-correction contribution (relax), frozen-density (frozen) \cite{d2002}, Krappe-Nix-Sirk (KNS) \cite{kns}, Bass (Bass80) \cite{bass80}, Bass (Bass74) \cite{bass73}, proximity (Prox77) \cite{prox77}, proximity (Prox2000) \cite{prox2000}, Winter \cite{winther}, and Siwek-Wilczynska-Wilczynski (SWW) \cite{sww} potentials.}
\end{center}
\end{figure}

\begin{figure}
\begin{center}
\includegraphics[width=8.7cm]{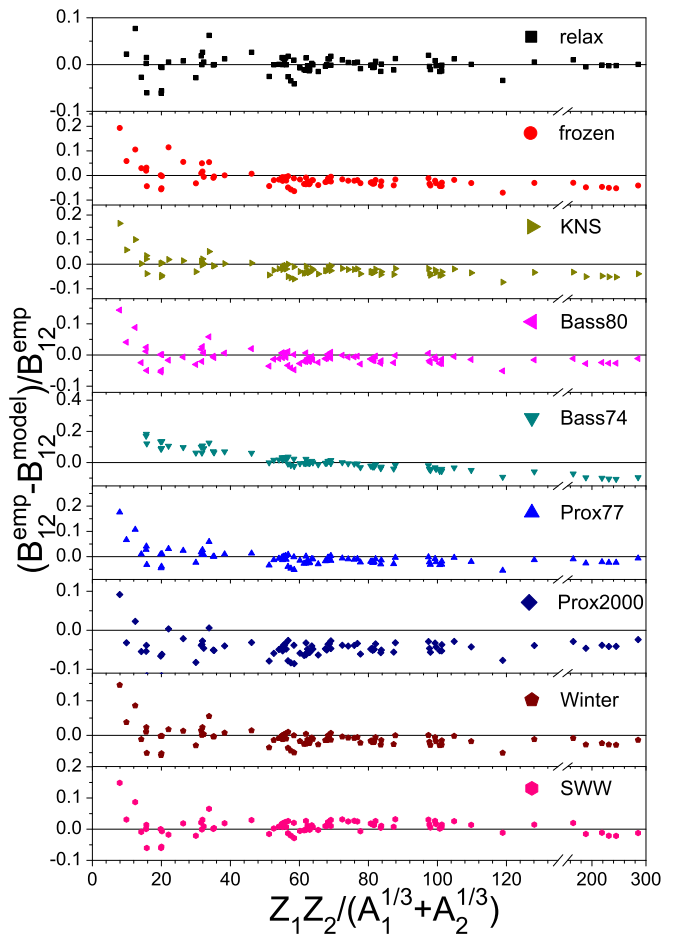}
\caption{(Color online) Relative differences between 89 empirical barrier heights and barrier heights evaluated with the help of various potentials. The notations are the same as in Fig. 1.}
\end{center}
\end{figure}

The values of 89 empirical barrier heights between spherical nuclei \cite{duttp,nbdhgmh,mintjw,gg} have been thoroughly described by a phenomenological relaxed-density potential with shell-correction contributions. The root mean square error of the barrier height description related to the second term of Eq. (19) is 0.879 MeV. The value of the same quantity obtained without the shell-correction contribution to the potential is 0.976 MeV. The comparison of the root mean square errors shows that the shell-correction contribution to the nucleus-nucleus potential is an important ingredient which improves the description of empirical barrier heights.

The root mean square errors related to the description of 89 empirical barrier heights obtained in the frameworks of the frozen-density \cite{d2002}, Krappe-Nix-Sirk \cite{kns}, Bass 1974 \cite{bass73}, Bass 1980 \cite{bass80}, proximity 1977 \cite{prox77}, proximity 2000 \cite{prox2000}, Winter \cite{winther}, and Siwek-Wilczynska-Wilczynski \cite{sww} potentials are 3.373, 3.460, 6.718, 1.815, 1.784, 3.710, 1.835, and 1.521 MeV, respectively. These values of the root mean square errors are larger than the one obtained using the phenomenological relaxed-density potential with shell-correction contributions.

The absolute differences between 89 empirical barrier heights and the barrier heights evaluated using various phenomenological potentials are presented in Fig. 1. As a rule, various empirical potentials well describe the values of the barrier heights for light nucleus-nucleus systems. The values of the barrier heights evaluated with the help of phenomenological potentials from Refs. \cite{bass80,bass73,kns,winther,prox2000,d2002} for heavy and especially very heavy nucleus-nucleus systems are larger than the empirical ones.
It is also useful to compare the relative differences between the empirical barrier heights and the barrier heights obtained using various phenomenological potentials. The corresponding differences are presented in Fig. 2. The relative differences between the empirical and theoretical barrier heights clearly show the accuracy of the barrier evaluation for light systems in the framework of various approaches.

The analysis of the results presented in Figs. 1 and 2 shows that phenomenological relaxed-density potential with shell-correction contributions successfully describes the empirical barrier heights in the full range of nucleus-nucleus systems in contrast to other potentials.

\begin{figure}
\begin{center}
\includegraphics[width=8.7cm]{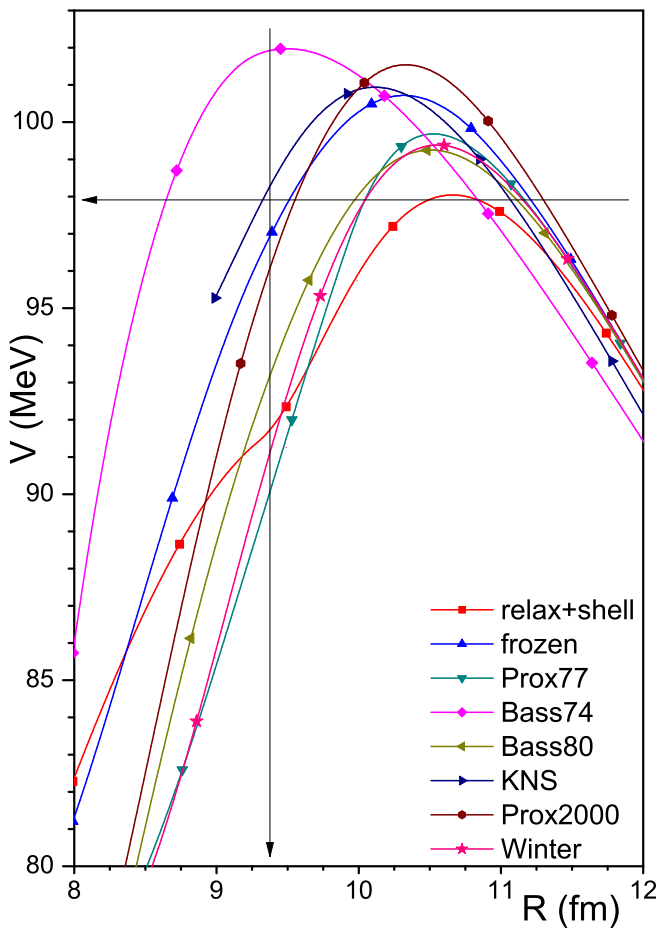}
\caption{(Color online) Nucleus-nucleus potentials for the system $^{58}$Ni+$^{58}$Ni evaluated with the help of various approaches. The empirical value of the barrier height is shown by a horizontal arrow and the touching distance of these nuclei by a vertical arrow. }
\end{center}
\end{figure}

The nucleus-nucleus potentials for the system $^{58}$Ni+$^{58}$Ni evaluated in various approaches are presented in Fig. 3. The value of the shell-correction energy for $^{58}$Ni is $- 4.45$ MeV. The barrier height of the phenomenological relaxed-density potential with the shell-correction contribution is close to the empirical value 97.90 MeV \cite{gg}; see Fig. 3. The relaxed-density potential deviates from other potentials around the touching distance and at smaller distances.

There are many nuclei which have positive values of the shell-correction energies. Therefore, it is interesting to study the effect of the sign of the shell-correction energy on the nucleus-nucleus potential. The full relaxed-density potentials obtained with opposite values $\mp$4.45 MeV of the shell-correction energy of $^{58}$Ni and the macroscopic part of the potential (the full potential without the shell-correction contribution) are presented for the system $^{58}$Ni+$^{58}$Ni in Fig. 4. The shell-correction contribution to the potential is small at large distances and is important around the touching point (see Fig. 4).

\begin{figure}
\begin{center}
\includegraphics[width=8.7cm]{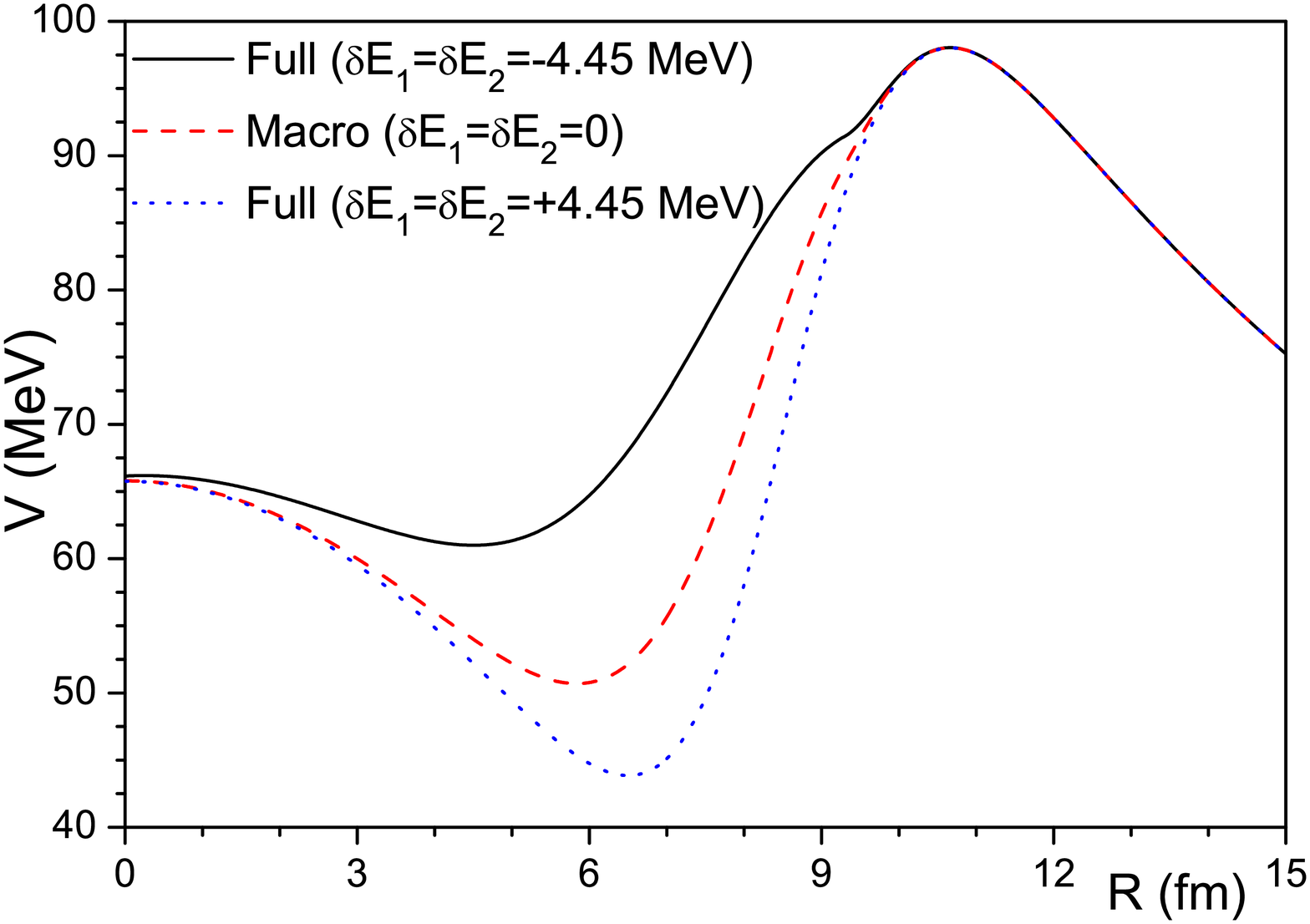}
\caption{(Color online) Full relaxed-density potential with the shell-correction contribution for the system $^{58}$Ni+$^{58}$Ni evaluated for various contributions of the shell-correction energies to the potential.}
\end{center}
\end{figure}

According to Eqs. (1) and (15) the depth of the capture well of the relaxed-density potential with negative shell-correction contributions (the case $\delta E_1 + \delta E_2<0$) becomes shallower than the one of the macroscopic potential. The barrier thickness of the full potential is larger than the one for the macroscopic potential. In contrast to this, the shell-correction contribution to the potential at $\delta E_1 + \delta E_2 > 0$ reduces the barrier thickness and increases the well depth of the full potential in comparison to the macroscopic ones. This effect is clearly seen in Fig. 4. Because of this effect the deep sub-barrier fusion hindrance takes place for a nucleus-nucleus system with the strong negative shell-correction contribution to the full heavy-ion potential, while the strong positive shell-correction contribution to the full potential leads to the weak enhancement of the deep sub-barrier fusion cross section; see Ref. \cite{dhindr} for details.

In conclusion, the phenomenological relaxed-density nucleus-nucleus potential with the shell-correction contribution is discussed in detail. The shell-correction contribution to the potential is related to inner structure of nuclei, which is disturbed by nucleon-nucleon interactions of colliding nuclei. A simple approach for the evaluation of the shell correction contribution to the full potential is proposed. The shell-correction contribution shows how the full potential for the specific nucleus-nucleus system deviates from the global macroscopic potential. The global macroscopic potential smoothly depends on $A_1$, $Z_1$, $A_2$ and $Z_2$ in contrast to the shell-correction part. The shell-correction contribution to the full potential is very important at distances smaller than the barrier radius. The phenomenological relaxed-density nucleus-nucleus potential with the shell-correction contribution can reproduce the empirical barrier heights with a value of the root mean square error of 0.879 MeV.

\section*{Acknowledgments}

The author thanks Professor A. V. Afanasjev (Mississippi State University), Professor M. I. Sitnov (the Johns Hopkins University Applied Physics Laboratory), and Professor V. I. Tretyak (Institute for Nuclear Research, Kiev) for useful remarks.

\end{document}